\documentclass[twocolumn,prl,aps,showpacs]{revtex4}

\usepackage{amssymb}
\usepackage{amsfonts}
\usepackage{amsmath}
\usepackage{epsfig}
\usepackage{graphicx}
\UseRawInputEncoding
\begin{document}

\title{

{Temporal characterization of electron dynamics in attosecond XUV and infrared laser fields}
\footnotetext{$^{*}$ guoli@shnu.edu.cn}
\footnotetext{$^{\ddag}$ chen$\_$jing@iapcm.ac.cn}
 }

\author{Li Guo$^{1,2*}$}
\author{Yi Jia$^{3}$}
\author{Mingqing, Liu$^{4}$}
\author{Xinyan Jia$^{3}$}
\author{Shilin Hu$^{5}$}
\author{Ronghua Lu$^{2}$}
\author{ShenSheng Han$^{2}$}
\author{Jing Chen$^{4,6,{\ddag}}$}

\affiliation{$^{1}$ Department of Physics, Shanghai Normal University, Shanghai 200234, China}

\affiliation{$^{2}$ Key laboratory for Quantum Optics, Shanghai Institute of Optics and Fine Mechanics,
Chinese Academy of Sciences, Shanghai 201800, China}
\affiliation{$^{3}$ Quantum Optoelectronics Laboratory, Southwest Jiaotong University, Chengdu 610031, China}
\affiliation{$^{4}$ Institute of Applied Physics and Computational Mathematics, P. O.
Box 8009, Beijing 100088, China}
\affiliation{$^{5}$  Guangdong Provincial Key Laboratory of Quantum Metrology and Sensing $\&$ School of Physics and Astronomy, Sun Yat-Sen University (Zhuhai Campus), Zhuhai 519082, China}
\affiliation{$^{6}$ Center for Advanced Material Diagnostic
Technology, College of Engineering Physics, Shenzhen Technology
University, Shenzhen 518118, China}


\begin{abstract}
We use a Wigner distribution-like function based on the strong field approximation theory to obtain the time-energy distributions and the ionization time distributions of electrons ionized by an XUV pulse alone and in the presence of an infrared (IR) pulse. In the case of a single XUV pulse, although the overall shape of the ionization time distribution resembles the XUV-envelope, its detail shows dependence on the emission direction of the electron and the carrier-envelope phase of the pulse, which mainly results from the low-energy interference structure. It is further found that the electron from the counter-rotating term plays an important role in the interference. In the case of the two-color pulse, both the time-energy distributions and the ionization time distributions change with varying IR field. Our analysis demonstrates that the IR field not only modifies the final electron kinetic energy but also changes the electron's emission time, which results from the change of the electric field induced by the IR pulse. Moreover, the ionization time distributions of the photoelectrons emitted from atoms with higher ionization energy are also given, which show less impact of the IR field on the electron dynamics.
\end{abstract}

 \maketitle
\section{Introduction}
The development of the ultrashort light pulse, e.g. isolated attosecond pulse, allows us to probe the electron dynamics of atoms and molecules which occur on the attosecond timescale \cite{mod,2012jpb,review3,pengliang}. The first experiment of an isolated XUV pulse has been reported by Hentschel \emph{et. al} \cite{2001-exp}. Temporal characterization of the attosecond XUV pulse can be achieved using the attosecond streaking technique \cite{asc,2001-exp,67} and FROG-CRAB (Frequency-resolved optical gating-complete reconstruction of attosecond bursts) scheme \cite{frog-c,15, 16}. Both methods need to measure the electron spectrum generated by the XUV pulse in the presence of a few-cycle infrared (IR) field which is a function of the relative phase between the XUV and the IR pulses. It is worthwhile mentioning that the properties of the IR field can also be determined in the same measurement \cite{lwave}.

On the other hand, the time of photoionization can be studied using an XUV pulse, which is achieved by combination of the XUV pulse and an IR field, such as one attoseccond XUV pump field and a phase-locked IR probe field \cite{44,45}. Schultze \emph{et al.} obtained a time delay of $\thicksim$ 21 attoseconds (as) in the emission of electron liberated from the 2$p$ orbital of neon atom with respect to that released from the 2$s$ orbital at an XUV photon energy of 100 eV \cite{45}. In most measurements, the probe IR fields are generally weak enough not to ionize atoms but strong enough to modify the final energy distribution of the photoelectron liberated by the XUV pulse. Strictly speaking, this photoionization process is laser-assisted, where the system is simultaneously pumped and probed. In other words, the electron is ionized by the superposition of the electric fields of the XUV and the IR pulses. The effect of the IR field on the electron dynamics cannot be neglected. For example, the coupling of the streaking field and long-range tail of the Coulomb potential can lead to the time delays \cite{2012jpb,RMP2,zhang1,Ivan1} and the streaking field can cause the polarization of the atom \cite{pola,Ivan1}. Very recently, Chen \emph{et al.} have investigated the time evolution of complex amplitudes of bound state dynamics during the interaction with an optical laser pulse using attosecond streaking spectroscopy \cite{Lupx}.

In this paper, we mainly investigate, from the angle of the temporal distribution of the photoemission rate, how the IR field causes the change of the photoemission time with varying relative phase between the XUV and IR fields and with varying IR intensity in the absence of the Coulomb potential. The Wigner distribution-like function (WDL) based on the strong field approximation (SFA) theory \cite{lpb-guo,oe-guo,pra-guo,pra2-guo} has been adopted to calculate the time-energy distributions and the ionization time distributions of electrons. We first study in detail the dynamics of electrons ionized by an XUV pulse alone and then add an IR field to investigate its effect on the dynamic characteristics of electrons compared to those in the XUV alone. It should be noted that in this paper relatively low or moderately IR intensity is considered, which is generally used in the conventional attosecond streaking experiments. When the IR intensity is strong, it can cause significant electron ionization. Hence, the rescattering electron can absorb an XUV photon when it revisits the core \cite{475} or can interfere with photoelectron of one XUV-photon ionization streaked by the strong IR field \cite{458,459}, which needs other theoretical models to be investigated and is beyond the scope of the present work.

\section{Theory method}
The definition and the derivation of the WDL function based on the SFA theory are given in more details in Refs. \cite{lpb-guo,oe-guo}. Here, for simplicity, we directly give the WDL function in one-dimensional system, which is given as follows:
\begin{equation}
f(t,\frac{p^{2}}{2},\Theta)=\frac{1}{\pi}\int_{-\infty}^{\infty}S''^{*}(t+t',\Theta)S''(t-t',\Theta)e^{-2i\frac{p^{2}}{2}t'}dt',\label{1}
\end{equation}
where $\frac{p^{2}}{2}$ is the final kinetic energy of the electron, $\Theta$ is the emission angle of the photoelectron with respect to a coordinate axis (the positive x axis in this paper), and $S''=\frac{S'}{\sqrt{p}}$. The variable $S'$ and the first term $S_{fi}$ (namely the SFA theory \cite{Reiss}) of the $S$-matrix expansion satisfy the relation as follows:
\begin{eqnarray}
S_{fi}&=&\frac{-i}{\sqrt{v}}\int_{-\infty}^{\infty}dt\left\langle\mathbf{p+A}(t)\right\vert-\mathbf{r}\cdot
\mathbf{E}(t)\left\vert\varphi_{0}\right\rangle
\nonumber\\
&&\times \exp\left[i\int_{-\infty}^{t}[\mathbf{p}\cdot \mathbf{A}(\tau)+\frac{A(\tau)^{2}}{2}]d\tau+iI_{p}t+i\frac{p^{2}}{2}t\right]\nonumber\\
&=&\frac{1}{\sqrt{2\pi}}\int_{-\infty}^{\infty}dt
S'e^{i\frac{p^{2}}{2}t},\label{2}
\end{eqnarray}
 where $\mathbf{A}(t)$ and $\mathbf{E}(t)=-\frac{\partial \mathbf{A}(t)}{\partial t}$ are the vector potential and the electric field of the pulse, respectively. $I_{p}$ is the ionization energy of atoms.

The WDL function satisfies the following condition, which can be proved using the same method as shown in Refs. \cite{lpb-guo,oe-guo}:
\begin{equation}
\int f(t,\frac{p^{2}}{2})dt=|S_{fi}|^{2}/p.\label{3}
\end{equation}
One can find that the above formula gives the same energy spectrum as calculated by the SFA in one-dimensional system.

Further, we can obtain the ionization time distribution (ITD) by integrating the WDL function over the energy.
\begin{equation}
P(t)=\int f(t,\frac{p^{2}}{2})d(\frac{p^{2}}{2}).\label{4}
\end{equation}

In this paper, the vector potential of the two-color field is the sum of $\mathbf{A}_{IR}(t)$ and $\mathbf{A}_{XUV}(t)$ which are:
\begin{eqnarray}
\textbf{A}_{IR}(t)&=&-\frac{E_{IR}}{\omega}\sin^{2}\left[\frac{\pi
t}{T_{IR}}\right]\cos(\omega t+\varphi_{IR})\textbf{e}_{x},\nonumber\\
&&0<t<T_{IR}\label{5}
\end{eqnarray} and
\begin{eqnarray}
\textbf{A}_{XUV}(t)&=&-\frac{E_{XUV}}{\Omega}\sin^{2}\left[\frac{\pi
(t-t_{C}+\frac{T_{XUV}}{2})}{T_{XUV}}\right]\nonumber\\
&&\times\cos[\Omega(t-t_{C}+\frac{T
_{XUV}}{2})+\varphi_{XUV}]\textbf{e}_{x}, \nonumber\\  &&t_{C}-\frac{T_{XUV}}{2}<t<t_{C}+\frac{T_{XUV}}{2}\label{6}
\end{eqnarray} respectively,
where $T_{IR}$ ($T_{XUV}$), $\varphi_{IR}$ ($\varphi_{XUV}$) and $\omega$ ($\Omega$) are the pulse duration, the carrier-envelope phase (CEP), and laser frequency of the IR (XUV) field, respectively. $t_{C}$ is the time corresponding to the center of the XUV pulse and $\textbf{e}_{x}$  the unit vector along the x axis.

In this paper, we mainly use a model atom with ionization energy $I_{p}=2$ a.u., which is ionized by a single XUV pulse alone and the two-color field. Atomic units (a.u.) are used in this paper unless otherwise indicated. The peak intensity ($1\times 10^{14}$ W/cm$^2$) and the frequency ($\Omega=3$ a.u.) of the XUV laser pulse are fixed.

\section{Results and discussions}
\subsection{A. Results in an XUV pulse}

\begin{figure}[tbh]
\begin{center}
\rotatebox{0 }{\resizebox *{9cm}{6cm} {\includegraphics
{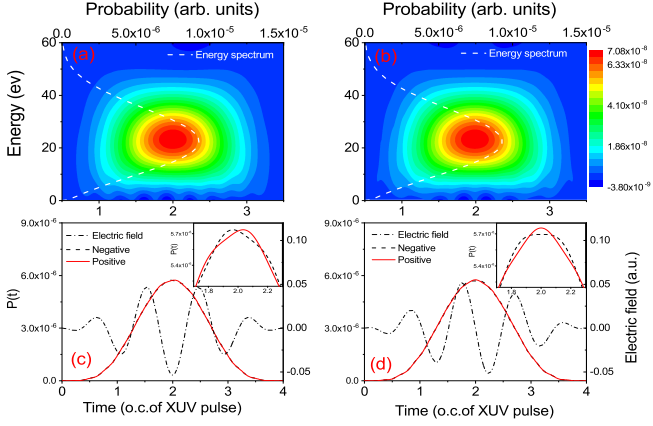}}}
\end{center}
\caption{The TEDs of electrons along the positive x axis ($\Theta=0$) emitted by a single XUV pulse with $\varphi_{XUV}=0.5\pi$ (a) and 0 (b). The white dashed lines in the panels (a) and (b) are the energy spectra calculated by the SFA. The ITDs of electrons for $\varphi_{XUV}$=0.5$\pi$ (c) and 0 (d). The insets in panels (c) and (d) denote the enlarged peaks of the ITDs. The corresponding electric field of the XUV pulse is also given (dashed dotted lines).}
\label{fig1}%
\end{figure}
\begin{figure}[tbh]
\begin{center}
\rotatebox{0 }{\resizebox *{9cm}{8.5cm} {\includegraphics
{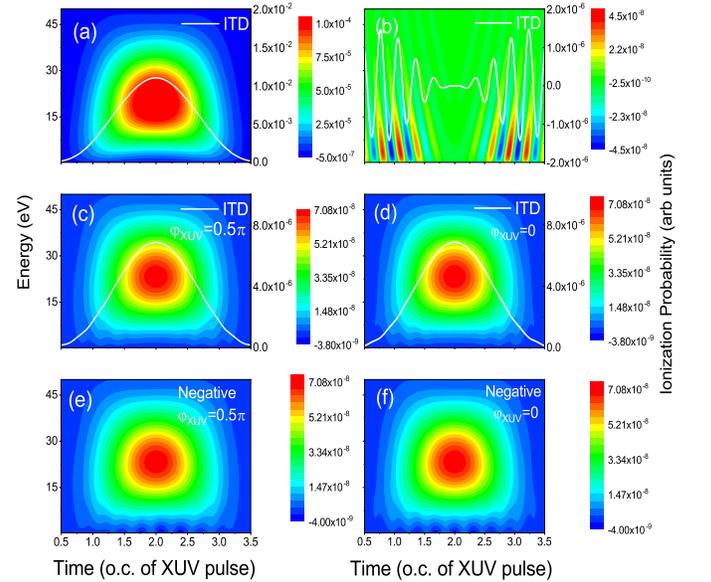}}}
\end{center}
\caption{The TEDs of electrons emitted by a single XUV pulse from the RWT component (a), the CRT component (b), and both two components for $\varphi_{XUV}=0.5\pi$ (c) and $\varphi_{XUV}=0$ (d). The dressing of the XUV field is neglected in the calculations of panels (a)-(d). Panels (e) and (f) show the TEDs of electrons which are the same as those of Figs. \ref{fig1}(a) and \ref{fig1}(b), respectively, but for the negative direction. The white line in each panel represents the ITD (see text). }
\label{fig2}%
\end{figure}

We first investigate the dynamic behaviors of electrons emitted only by a single XUV pulse with $t_{C}=\frac{T_{XUV}}{2}$. Figures \ref{fig1}(a) and \ref{fig1}(b) show the time-energy distributions (TEDs) of electrons along the positive x axis (namely $\Theta=$0, called as "positive direction" below) in a four-cycle XUV pulse with $\varphi_{XUV}$=$0.5\pi$ and 0, respectively, which are calculated by Eq. (\ref{1}). Noted that in order to intuitively show the physical meaning of the WDL function, the time-energy distribution is used to describe the result obtained by Eq. (\ref{1}). The energy spectra (white dashed lines) calculated by the SFA theory are also given in Figs. \ref{fig1}(a) and \ref{fig1}(b). One can find that both TEDs look the same and exhibit a broad distribution without fine structures presented in the low-frequency case (See Fig. 2 of Ref. \cite{pra-guo}). The maxima of the TEDs are at times close to the time of the center of the XUV pulse and at energies corresponding to the peaks of the energy spectra.
In addition, the TEDs in the negative direction ($\Theta=\pi$) also look like those shown in Figs. \ref{fig1}(a) and \ref{fig1}(b), which are plotted in Figs. \ref{fig2}(e) and \ref{fig2}(f).
Further, we integrate the TED over the energy to obtain the ITD in Figs. \ref{fig1}(c) and \ref{fig1}(d).
 As shown in Figs. \ref{fig1}(c) and \ref{fig1}(d), all ITDs are in the shape of the XUV envelope, which are consistent with the previous understanding that the temporal profile of the XUV-induced photoemission rate follows the intensity profile of the incident XUV pulse \cite{mod}. It is important to point out that in the TEDs there are many interference stripes located in the low-energy region and their positions vary with the CEP of the XUV pulse and the electron's emission direction.

 Give a closer glance at the peaks of the ITDs which are zoomed into in the insets of Figs. \ref{fig1}(c) and \ref{fig1}(d). One can find that minor differences exist among these ITDs. For $\varphi_{XUV}=0.5\pi$, the peak positions of the ITDs in both negative and positive directions slightly deviate from the center of the XUV pulse. There is a left shift in the positive direction with respect to the center of the XUV pulse while a right shift in the negative direction. The absolute values of left and right shifts are equal and estimated to be less than 1.68 as. As shown in the inset of Fig. \ref{fig1}(c), we can find that the ITDs in the positive and negative directions are mirror symmetric with respect to the central axis of the XUV pulse. For $\varphi_{XUV}$=0, the ITD is symmetric with respect to the center of the XUV pulse, in regardless of the emission direction of the electron. However, the maximal value of the ITD in the positive direction is a bit higher than that in the negative direction.

  All these fine differences shown in Figs. \ref{fig1}(c) and \ref{fig1}(d) indicate that for a single-photon ionization event induced by an XUV pulse, the ionization characteristic of the electron still has a weak dependence on the emission direction of the electron and the CEP of the XUV pulse. According to our calculations, it is the interference appearing in the low-energy region of the TED that is responsible for the dependence mentioned above.

Next, we investigate the origin of these interference structures.
 We can approximately write the XUV electric field $\mathbf{E_{XUV}}(t)$ as $(\mathbf{E_{-}}+\mathbf{E_{+}})/2$, where $\mathbf{E_{\mp}}=E_{XUV} f_{0}(t)e^{\mp i(\Omega t+\varphi_{XUV}-\frac{\pi}{2})}\textbf{e}_{x}$ and the slowly varying pulse envelope is assumed, i.e. $f_{0}(t)=\sin^{2}\left[\frac{\pi t}{T_{XUV}}\right]$. When $\mathbf{E_{-}}$ is used, the term $e^{i(\frac{p^2}{2}+I_{p}-\Omega)t}$ is present in the transition matrix element, which corresponds to the rotating wave term (RWT). This transition process satisfies energy conservation, in which an electron transitions from the bound state to the continuum state by absorbing an XUV photon. When $\mathbf{E_+}$ is adopted, the term $e^{i(\frac{p^2}{2}+I_{p}+\Omega)t}$ appears in the transition matrix element, which corresponds to the counter-rotating term (CRT). This ionization process does not satisfy energy conservation where an XUV photon is emitted during the electron transition.

The TEDs of the electrons calculated by the RWT and the CRT components are given in Figs. \ref{fig2}(a) and \ref{fig2}(b), respectively, which show no dependence on the electron's emission direction or the CEP of the pulse. The white lines in Fig. \ref{fig2} denote the ITD. As shown in Fig. \ref{fig2}(a), the RWT component dominates the distribution and no interference exists in the TED or ITD. In contrast, for the results from the CRT component shown in Fig. \ref{fig2}(b), there are many strong interference structures in the TED and the corresponding ITD (white line) oscillates. The contribution from the CRT component is much smaller than that from the RWT component (See the color bars of Figs. \ref{fig2}(a) and \ref{fig2}(b)). When both two components are considered (namely, $\mathbf{E_{XUV}}(t)=-\frac{\partial \mathbf{A_{XUV}}(t)}{\partial t}$ is used), a very weak interference structure located in the low-energy region shows up in the TEDs which are displayed in Fig. \ref{fig2}(c) for $\varphi_{XUV}=0.5\pi$ and in Fig. \ref{fig2}(d) for $\varphi_{XUV}=0$. The weak interference structure shows a dependence on $\varphi_{XUV}$ by comparison of the interference patterns shown in Figs. \ref{fig2}(c) and \ref{fig2}(d). This indicates that the CRT component not only survives but also interferes with the RWT component although its contribution is relatively small. It is worth noting that the interference between the RWT and CRT components can also be found in Ref. \cite{Lupx}, which leads to a weak oscillation in the time-dependent population of the bound state in the attosecond streaking field. Up to now, the structures in Figs. \ref{fig2}(a)-(d) have a common feature that they are symmetric with the center of the XUV pulse and are independent of the electron's emission direction.
It should be noted that we neglect the dressing of the XUV field (namely $A_{XUV}(t)=0$) in the calculations of Figs. \ref{fig2}(a)-\ref{fig2}(d). This is done for one reason that we can check the effect of the variable $A_{XUV}(t)$ on the electron dynamics because this variable is usually neglected in the case of an XUV (XUV train) field combined with an IR field due to its small value, e.g., $A_{XUV}(t)\sim0.01$ a.u. in our paper.

When the variable $A_{XUV}$ is considered, the interference stripes located in the low-energy region which are shown in Figs. \ref{fig2}(e) and \ref{fig2}(f) become clearer compared with those shown in Figs. \ref{fig2}(c) and  \ref{fig2}(d) and exhibit the dependence on the electron's emission direction. This demonstrates that although the value of $A_{XUV}(t)$ is very small, it can still affect the dynamics of the electron. Figure. \ref{fig2}(e) (Figure. \ref{fig2}(f)) shows the TED of the electron emitted along the negative direction in an XUV field with the same parameters as in Fig. \ref{fig1}(a) (Fig. \ref{fig1}(b)), except for $\Theta=\pi$. For $\varphi_{XUV}=0$, the constructive (destructive) interference occurs at the central time of the XUV pulse in the positive (negative) direction, which leads to a sharp (flat) peak of the ITD shown in Fig. \ref{fig1}(d). For $\varphi_{XUV}=0.5\pi$, the interference structure is no longer symmetric with respect to the center of the XUV envelope. The position of the constructive interference stripe closest to the peak of the envelope in the positive (negative) direction that is shown in Fig. \ref{fig1}(a) (Fig. \ref{fig2}(e)) is at the right (left) side of the peak of the XUV envelope, which makes a right (left) shift of the peak of the ITD in positive (negative) direction.

Based on the above calculations, it is concluded that the interference structure presenting in the low-energy region comes from the interference between the RWT and CRT components, which is responsible for the dependence of the ITD on $\varphi_{XUV}$ and the electron's emission direction. It is worth mentioning that the ratio of the transition amplitude of the electrons from the RWT component to that from the CRT components determines the strength of the interference structure in the TED and thus affects the ITD. According to our analysis, for a fixed XUV frequency, the larger the value of $I_p$, the weaker the effect of the interference on the ITD is.

\subsection{B. Results in a two-color pulse }
 In this section, we add an IR field to investigate its effect on the dynamical characteristics of electrons compared with the case of only a single XUV pulse. In this paper, we choose the IR pulse with fixed $\varphi_{IR}$=0 and a duration of 4 o.c. All times in the following text are in units of optical cycle (o.c.) of the involved IR field. Note that we use the arrival time (namely $t_{C}$) of the center of the XUV pulse in the presence of the IR field instead of the relative phase between the two fields to describe one specific two-color field.
 \begin{figure}[tbh]
\begin{center}
\rotatebox{0 }{\resizebox *{8cm}{8.5cm} {\includegraphics
{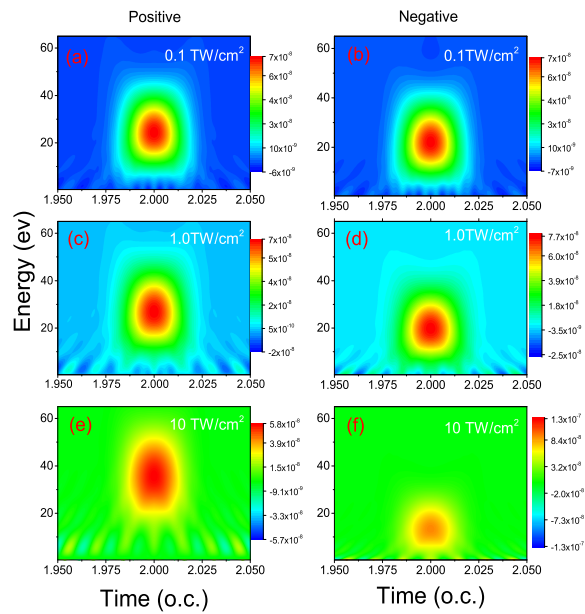}}}
\end{center}
\caption{The TEDs of attosecond streaking with $t_{C}=2.0$ o.c. in the positive (left column) and negative (right column) directions. The intensities of the IR fields are $1\times10^{11}$ W/cm$^{2}$ ((a) and (b)), $1\times10^{12}$ W/cm$^{2}$ ((c) and (d)) and $1\times10^{13}$ W/cm$^{2}$  ((e) and (f)), respectively. Here, $\varphi_{XUV}=0$. }
\label{fig3}%
\end{figure}

First, we take the case of attosecond streaking with $t_{C}=2$ o.c. and $\varphi_{XUV}=0$ as an example.
Figure \ref{fig3} shows the TEDs of attosecond streaking in the positive (left column) and negative (right column) directions for three IR intensities of $I_{IR}=1\times10^{11}$ W/cm$^{2}$ (Figs. \ref{fig3}(a) and \ref{fig3}(b)), $1\times10^{12}$ W/cm$^{2}$ (Figs. \ref{fig3}(c) and \ref{fig3}(d)), and $1\times10^{13}$ W/cm$^{2}$ (Figs. \ref{fig3}(e) and \ref{fig3}(f)).
As shown in Fig. \ref{fig3}, for the lowest intensity of $I_{IR}=1\times10^{11}$ W/cm$^{2}$, the TEDs in both the positive and negative directions are almost identical, which look the same as those of Fig. \ref{fig1}(b). However, when the IR intensity increases, the TEDs in the positive direction gradually shift to the higher energy region and the widths of the distributions become broader. In contrast, the TEDs in the negative direction have an opposite tendency. One can find that the difference of the TEDs between positive and negative directions is less pronounced for the lowest IR intensity ($I_{IR}=1\times10^{11}$ W/cm$^{2}$) and becomes more obvious for the higher IR intensities ($I_{IR}=1\times10^{12}$ W/cm$^{2}$ and $1\times10^{13}$ W/cm$^{2}$).

 These features can be approximately explained by a semiclassical picture as follows \cite{mod,2012jpb,pengliang,review3}: An electron in the bound state is ionized by absorbing an XUV photon to one continuum state with momentum $\mathbf{p}_{0}$ at time $t_{0}$, and then it moves only in the IR field. The final kinetic energy detected by the detector is given by
\begin{equation}
 \frac{[\mathbf{p}_{0}-\mathbf{A_{IR}}(t_{0})]^2}{2}\approx \frac{\mathbf{p}_{0}^2}{2}-\mathbf{p}_{0}\cdot\mathbf{A_{IR}}(t_{0}). \label{eq1}
 \end{equation}
 For the ionization time range shown in Fig. \ref{fig3}, the direction of the electron's drift momentum $-\mathbf{A_{IR}}(t_0)$ points to the positive direction, which leads to a high-energy (low-energy) shift of the photoelectron spectrum in the positive (negative) direction because of the positive (negative) value of $-\mathbf{p}_{0}\cdot\mathbf{A_{IR}}(t_{0})$.  Meanwhile, as shown in Fig. \ref{fig1}(a), for a given ionization time $t_{0}$, the initial momentum $p_{0}$ has a distribution with the value of the center slightly less than $\sqrt{2(\Omega-I_{p})}$. According to Eq. (\ref{eq1}), the value of the energy shift is proportional to $p_{0}$. This, together with the distribution of $p_{0}$, results in the change of the width of the TED. It is easy to understand that the discrepancy of the TEDs between these two directions is more obvious for the case of a high IR intensity because of the larger absolute value of the vector potential $\mathbf{A_{IR}}(t_0)$.
\begin{figure}[tbh]
\begin{center}
\rotatebox{0 }{\resizebox *{8cm}{12.cm} {\includegraphics
{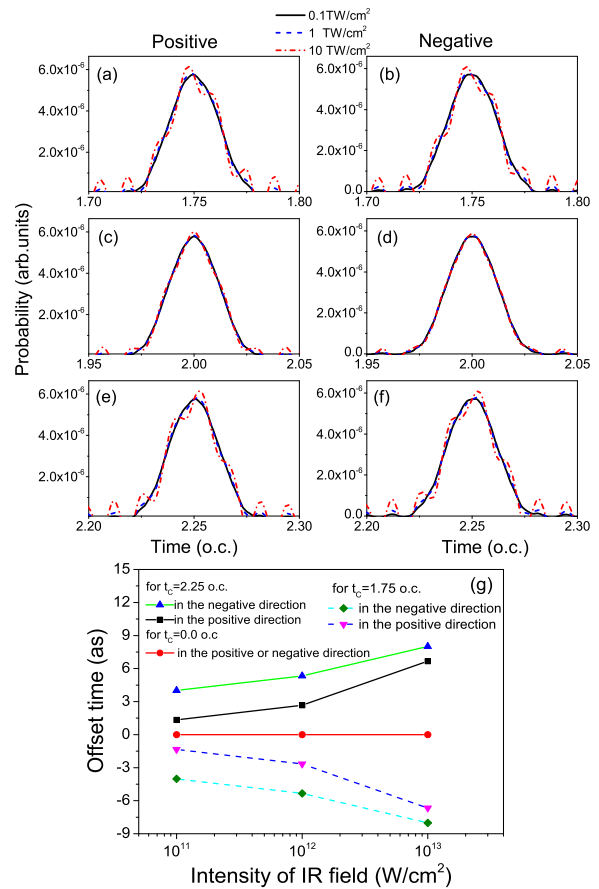}}}
\end{center}
\caption{The ITDs of attosecond streaking for three different cases in the positive ((a), (c), and (e)) and negative ((b), (d), and (f)) directions. The three cases are $t_{C}=$1.75 o.c. ((a) and (b)), 2.0 o.c. ((c) and (d)), and 2.25 o.c. ((e) and (f)), respectively. Panel (g) exhibits the offset time extracted from (a)-(f).}
\label{fig4}%
\end{figure}

 In order to show in detail the influence of the IR intensity on the ITD, we calculate the ITDs for three cases of $t_{C}=1.75$ o.c., 2.0 o.c., and 2.25 o.c. As shown in Figs. \ref{fig4}(a)-\ref{fig4}(f), the shapes of the ITDs in the same direction vary with different two-color fields. For the cases of $t_{C}=1.75$ o.c. and 2.25 o.c., the ITDs are no longer symmetric with respect to the center of the XUV pulse, the peaks of which obviously deviate from the center of the XUV pulse. However, for the case of $t_{C}=2.0$ o.c., the ITDs still show symmetric shapes with maxima at $t=t_{C}$. For clarification, the time offsets between the peak of the ITD and the center of the XUV pulse are plotted in Fig. \ref{fig4}(g) as a function of the IR intensity, which are extracted from Figs. \ref{fig4}(a)-\ref{fig4}(f). Here, the positive (negative) value of the time offset means time delay (advance). Several characteristics in the cases of $t_{C}=1.75$ o.c. and $t_{C}=2.25$ o.c. can be found in Fig. \ref{fig4}(g): (i) There are time delays in the case of $t_{C}=2.25$ o.c. while time advances in the case of $t_{C}=1.75$ o.c. (ii) The offset value becomes larger with increasing IR field for one specific case. (iii) For the electrons emitted along the same direction, the absolute value of the time offset in the case of $t_{C}=1.75$ o.c. is equal to that in the case of $t_{C}=2.25$ o.c. for the same IR intensity. (iiii) The time offset value of the ITDs in both positive and negative directions are unequal for a given case, which indicates that the weak dependence of the electron behavior on the emission direction.


\begin{figure}[tbh]
\begin{center}
\rotatebox{0 }{\resizebox *{7cm}{9.5cm} {\includegraphics
{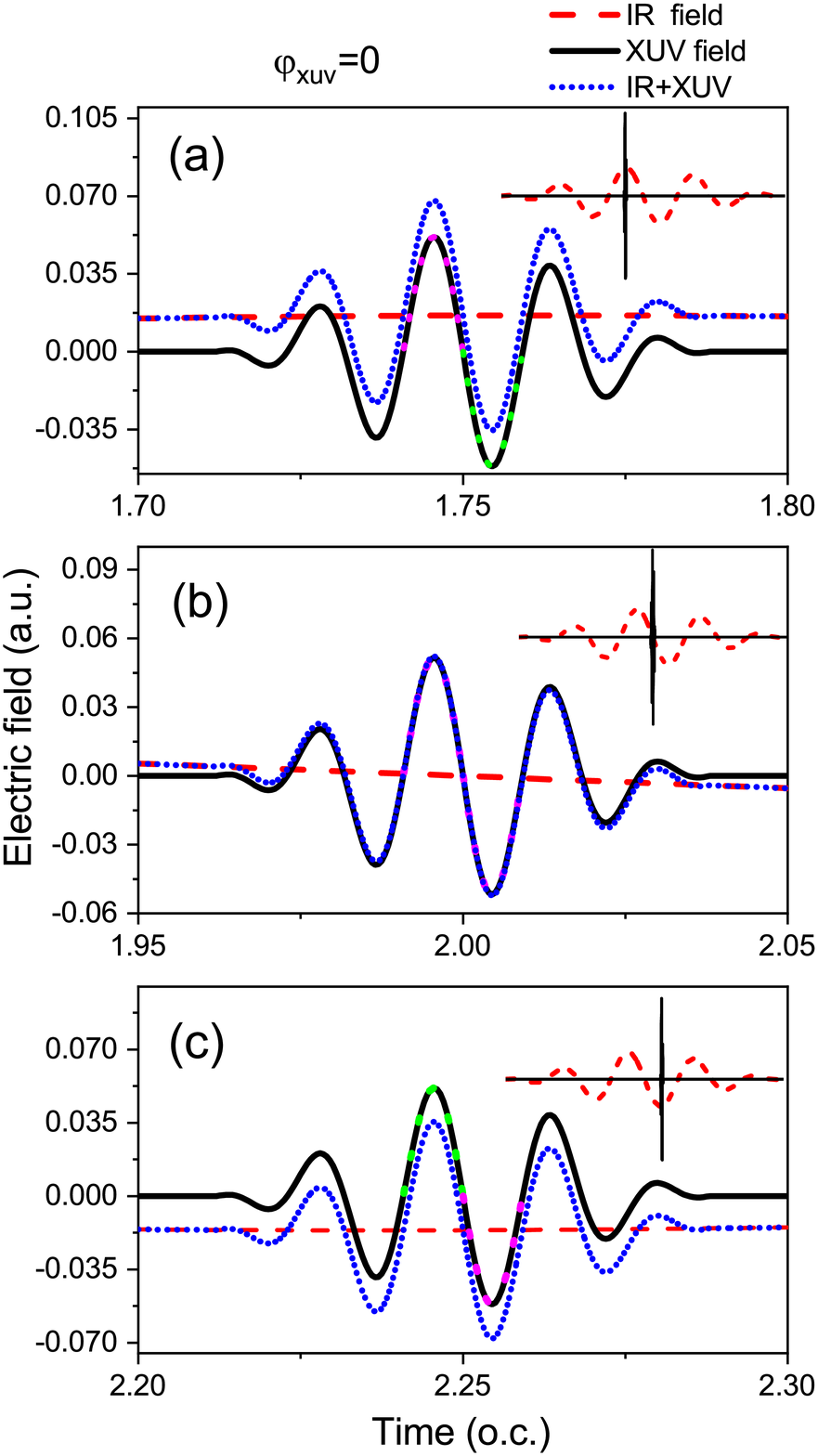}}}
\end{center}
\caption{The electric fields of the IR (red dashed lines) and the XUV (solid lines) pulses for the cases of $t_{C}=1.75$ o.c. (a), 2.0 o.c. (b), and 2.25 o.c. (c). Magenta (green) dotted lines denote the electric field of the XUV pulse which is enhanced (suppressed) after adding the IR field. The blue short dotted lines are the sum of the IR and XUV fields. Here $I_{IR}=1\times10^{13}$ W/cm$^{2}$. The whole two-color pulse is given at the top right of each panel.}
\label{fig5}%
\end{figure}

 These features can be ascribed to the change of the electric field. Figure \ref{fig5} shows the electric fields of the XUV and the IR pulses for three different cases. The whole two-color pulse is also given at the top right of each panel. The presence of the IR field is equivalent to the increase or the decrease in the original XUV field at times marked by the magenta dotted lines for the former and by the green dotted lines for the latter. Here, we mainly focus on the central one optical cycle of the XUV pulse, because the change of the ITD induced by the IR field is mainly confined to this cycle. For instance, the XUV field is enhanced at times $t<t_{C}$ and is suppressed at times $t>t_{C}$ in the case of $t_{C}=$1.75 o.c. This causes $\mathbf E(t)\neq -\mathbf E(2t_{C}-t)$ in the time range during the XUV pulse and thus leads to the unsymmetric shape of the ITD. Meanwhile, the peak of the ITD shifts to the side of the electric field enhancement. The similar analyses can be used in the case of $t_{C}=2.25$ o.c. One can find that the peaks of the ITDs in the case of $t_{C}=2.25$ o.c. also shift to the right side of the center of the XUV pulse where the electric field is enhanced.

 \begin{figure}[tbh]
\begin{center}
\rotatebox{0 }{\resizebox *{8.5cm}{10cm} {\includegraphics
{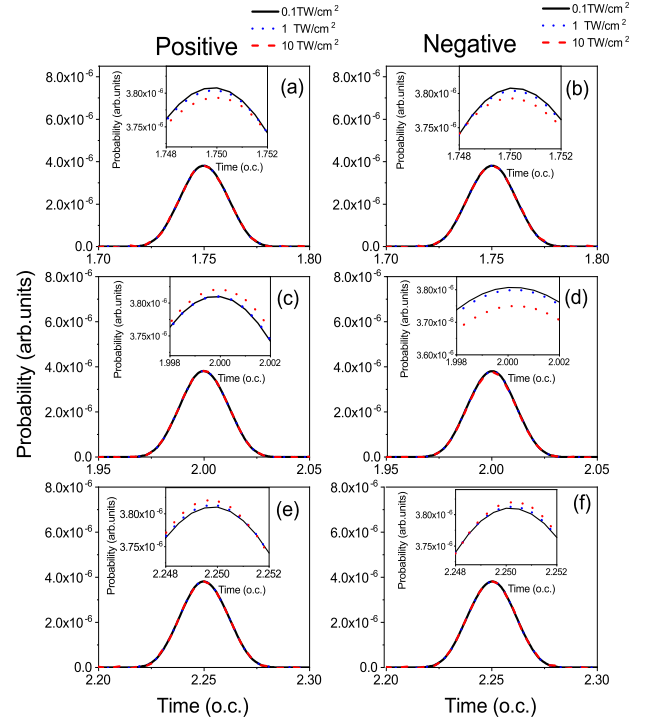}}}
\end{center}
\caption{The ITDs of attosecond streaking for three cases of $t_{C}=1.75$ o.c. ((a) and (b)), 2.0 o.c. ((c) and (d)), and 2.25 o.c. ((e) and (f)). Same as Fig. \ref{fig4} but for $\varphi_{XUV}=0.5\pi$ and $I_{p}=2.5$ a.u. The peak of the corresponding ITD is enlarged in the inset.}
\label{fig6}%
\end{figure}
 It is worth noting that there is a relationship of $\mathbf{E_{1}}(t_{C1}-\delta t)=-\mathbf{E_{2}}(t_{C2}+\delta t)$, where $\mathbf{E_{1}}$ ($\mathbf{E_{2}}$ ) and $t_{C1}$ ($t_{C2}$) represent the two-color field and the central time of the XUV pulse in the case of $t_{C}=1.75$ o.c. (2.25 o.c.), namely $t_{C1}=1.75$ o.c. ($t_{C2}=2.25$ o.c.), as shown in Fig. \ref{fig5}(a) (Fig. \ref{fig5}(c)), respectively, and $\delta t$ is the time difference between any time $t$ and the central time of the XUV pulse, ranging from -$\frac{\omega}{\Omega}$ o.c. to $\frac{\omega}{\Omega}$ o.c. in the present work. This characteristic of the electric fields results in the equal absolute values of the time offsets for these two cases under the same parameters. However, for the case of $t_{C}=2.0$ o.c., although the electric fields of the central one cycle of the XUV pulse are all increased (They are not visually obvious), they still satisfy the relation of $\mathbf E(t)=-\mathbf E(2t_{C}-t)$ (See Fig. \ref{fig5}(b)), leading to the symmetric shape of the ITDs.

 Based on the characteristics shown in Figs. \ref{fig3} and \ref{fig4}, we can find that the IR field not only causes the energy shift of the electron explained approximately by the semiclassical picture but also has a obvious effect on the emission time of the electron. This effect becomes greater with increasing IR intensity and more importantly, the time offsets induced by the IR field are different for different two-color fields.

 \begin{figure}[tbh]
\begin{center}
\rotatebox{0 }{\resizebox *{7.5cm}{9.5cm} {\includegraphics
{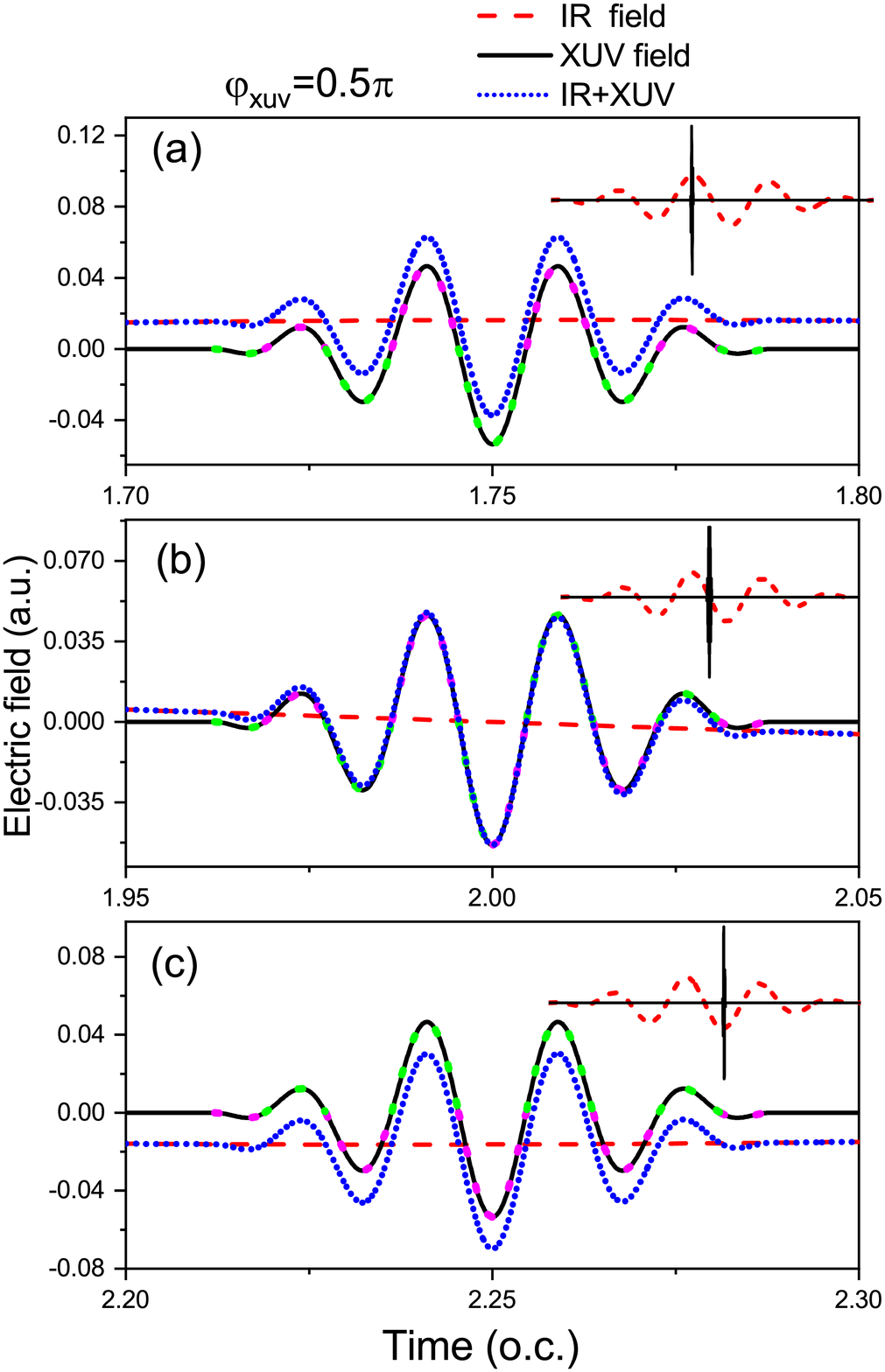}}}
\end{center}
\caption{The two-color fields for three different cases. Same as Fig. \ref{fig5} except for $\varphi_{XUV}=0.5\pi$. }
\label{fig7}%
\end{figure}

 Further, we study the effect of the IR field on the ITDs of electrons emitted from atoms with the higher ionization energy of $I_{p}=2.5$ a.u. Figure \ref{fig6} displays the ITDs of attosecond streaking for three different cases of $t_{C}=1.75$ o.c. (Figs. \ref{fig6}(a) and \ref{fig6}(b)), 2.0 o.c. (Figs. \ref{fig6}(c) and \ref{fig6}(d)), and 2.25 o.c. (Figs. \ref{fig6}(e) and \ref{fig6}(f)). The parameters are the same as in Fig. \ref{fig4}, except for $\varphi_{XUV}=0.5\pi$ and $I_{p}=2.5$ a.u. As can be clearly shown in Fig. \ref{fig6}, all ITDs are smooth curves and look almost identical. These curves still resemble the envelope of the XUV pulse. By comparing with those shown in Fig. \ref{fig4}, it is found that the impact of the IR field on the ITD is smaller for the case of $I_{p}=2.5$ a.u. Meanwhile, we zoom into the peak of the ITD in the inset of each panel. One can find that some tiny differences still exist among these ITDs. For each case, the ITDs in the positive and negative directions are also mirror symmetric with respect to the axis $t=t_{C}$, which is consistent with the case of only one XUV pulse with $\varphi_{XUV}=0.5\pi$ shown in Fig. \ref{fig1}(c). Another feature is that the ionization rates at ionization times in vicinity of the center of the XUV pulse, in regardless of the emission directions, increase in the case of $t_{C}=1.75$ o.c. while they decrease in the case of $t_{C}=2.25$ o.c., with increasing IR intensity. These are also the results of the change of the electric field (See Fig. \ref{fig7}). Figure \ref{fig7} displays the two-color pulse with the same parameter as in Fig. \ref{fig5}, except for $\varphi_{XUV}=0.5\pi$. As shown in Fig. \ref{fig7}, the pattern of the XUV field with $\varphi_{XUV}=0.5\pi$ changes with respect to that with $\varphi_{XUV}=0$ shown in Fig. \ref{fig5}.
  In the case of $t_{C}=1.75$ o.c., the electric field of the central half optical cycle of the XUV pulse is decreased and thus causes the suppression of the yield of the electron. However, the corresponding electric field is increased in the case of $t_{C}=2.25$ o.c., leading to the enhancement of the ionization rate. 

For the case of $t_{C}=2$ o.c., the situation is different. As can be clearly shown in insets of Figs. \ref{fig6}(c) and \ref{fig6}(d), with increasing IR intensity the ionization rates gradually increase for the electrons along the positive direction while they decrease for the electrons along the negative direction. This can be understood as follows: According to the semiclassical picture, for the electrons emitted along the negative direction, the direction of the initial momentum $\mathbf{p_0}$ after ionized by absorbing an XUV-photon is opposite to the direction of $-\mathbf{A_{IR}}(t_{0})$ under consideration. For the electrons with the small value of $p_0$, the original emission direction of the electron can be reversed by the IR field because of the larger value of $-\mathbf{A_{IR}}(t_{0})$, namely $[\mathbf{p}_{0}-\mathbf{A_{IR}}(t_{0})]<0$. More electrons meet above condition compared with the case of the low ionization energy $I_{p}=2$ a.u. This results in the lower yield of electron in the negative direction and the higher yield of electron in the positive direction, which is more obvious for the case of $I_{IR}=1\times10^{13}$ W/cm$^{2}$.

\section{Conclusion}
In conclusion, the WDL function based on the SFA theory is used to calculate the TED and ITD of electrons emitted by a single XUV pulse alone and two-color pulses with different IR intensities. For the case of an XUV pulse alone, the shapes of the ITDs with different calculation parameters all resemble the XUV-envelope. However, there are existing some fine differences in details which show the dependence of the ITD on the CEP of the XUV pulse and the electron's emission direction. The interference structure in low-energy region of the TED is mainly responsible for the above dependence. It is further found that the electron from the counter-rotating term plays an important role in the low-energy interference structure.

 For the case of the two-color pulses, both the TEDs and the ITDs vary with respect to those induced only by the XUV pulse. The TEDs clearly demonstrate that the IR field can cause the electron's energy shift and change the width of the electron's distributions, which can be explained approximately by the simple man picture that the electron ionized by absorbing one XUV photon is streaked by the IR field. Meanwhile, it is found that the peak positions of the ionization time distributions deviate from those generated only by the XUV pulse and shift to the side of the electric field enhancement, indicating the effect of the IR field on the electron's emission time. These effects are more obvious with increasing IR intensity and are different for the different delays between the XUV and IR fields, which can be attributed to the IR field-induced change of electric fields. Further, the ITDs of the photoelectrons emitted from atoms with the higher ionization energy are also given, on which the influence of the IR field is less pronounced. This implies that the impact of the IR field on the emission time of the electrons emitted from different initial states may also be different.

\section{Acknowledgments}
This work was partially supported by the National Key program for
S\&T Research and Development (Grants No. 2019YFA0307700 and No. 2016YFA0401100), the NNSFC (Grants No.
11774361, 11775286, and 11804405).


\end{document}